\pdfoutput=1
\documentclass{article}
\usepackage{spconf,amsmath,graphicx}

\usepackage{comment}
\usepackage{hyperref}
\usepackage{caption}

\usepackage{amssymb}
\usepackage{multirow} 
\graphicspath{{grf/}}

\newcommand{\calL}{\ensuremath{\mathcal{L}}}

\newcommand{\norm}[1]{\left\lVert#1\right\rVert}

\newcommand{\caja}[4][1]{{%
    \renewcommand{\arraystretch}{#1}%
    \begin{tabular}[#2]{@{}#3@{}}%
      #4%
    \end{tabular}%
    }}

{%
\begin{list}{#1}{
\vspace{-\topsep}
\vspace{-\partopsep}
\setlength{\itemindent}{0cm}
\setlength{\rightmargin}{0cm}
\setlength{\listparindent}{0cm}
\settowidth{\labelwidth}{#1}
\setlength{\leftmargin}{\labelwidth}
\addtolength{\leftmargin}{\labelsep}
\setlength{\itemsep}{0cm}
}%
}%
{%
\end{list}
\vspace{-\topsep}
\vspace{-\partopsep}
}

{\begin{enumerate}%
}%
{\end{enumerate}}

\usepackage{color}
\usepackage{colortbl}
\definecolor{light-gray}{gray}{0.8}

\newcommand{\karen}[1]{\textcolor{green}{#1 (Karen)}}

\renewcommand{\karen}[1]{}

\title{UNSUPERVISED PRE-TRAINING OF BIDIRECTIONAL SPEECH ENCODERS \\ VIA MASKED RECONSTRUCTION}
\name{Weiran Wang$^{\star}\thanks{Work done while Weiran Wang was at Amazon Alexa.}$ \qquad Qingming Tang$^{\dagger}$ \qquad Karen Livescu$^{\dagger}$}
\address{$^{\star}$Salesforce Research $\qquad$ $^{\dagger}$ Toyota Technological Institute at Chicago \\
\texttt{weiran.wang@salesforce.com $\qquad$ \{qmtang,klivescu\}@ttic.edu}}

\begin{document}
\maketitle

\begin{abstract}
We propose an approach for pre-training speech representations via a
masked reconstruction loss. Our pre-trained encoder networks are
bidirectional and can therefore be used directly in typical bidirectional
speech recognition models. The pre-trained networks can then be fine-tuned
on a smaller amount of supervised data for speech recognition.
Experiments with this approach on the LibriSpeech and Wall Street Journal
corpora show promising results.
We find that the main factors that lead to speech recognition improvements are: masking segments of sufficient width in both time and frequency, pre-training on a much larger amount of unlabeled data than the labeled data, and domain adaptation when the unlabeled and labeled data come from different domains.
The gain from pre-training is additive to that of supervised data
augmentation.
\end{abstract}

\vspace{-.05in}
\section{Introduction}
\label{s:intro}
\vspace*{-1.5ex}

We study the problem of improving speech recognition via unsupervised pre-training, possibly on external data.
Unsupervised pre-training has a long history in the field of speech recognition.
Restricted Boltzmann Machines (RBMs)~\cite{hinton2002training} were widely used to pre-train deep neural networks as part of a speech recognizer~\cite{Hinton_12a}, often on the same transcribed data used for acoustic modeling. 
\karen{was RBM-based pretraining ever used in the same setting as ours, i.e. pretraining on external data?}
In recent years, however, RBM-based pre-training has been largely abandoned, \karen{I removed the phrase about RBM-based pre-trained being slow, as the sentence is long and the other reasons are more important.  Feel free to bring it back if you think it is important.}
because direct supervised training of deep neural networks
has improved due to new techniques such as
better initialization~\cite{glorot2010understanding}, 
non-saturating activation functions~\cite{NairHinton10a}, and better control of generalization~\cite{Srivas_14a}.
However, very recent work has begun to reconsider the value of unsupervised pre-training, specifically in the context of representation learning on a large set of unlabeled data, for use in supervised training on a smaller set of labeled data~\cite{pascual2019learning,chung2019unsupervised,schneider2019wav2vec}. \par

At the same time, in 
the area of natural language processing (NLP), unsupervised pre-trained representation learning has been extremeley successful.  In the past two years, several approaches have been proposed for pre-trained text representations~\cite{peters2018deep, lagler2013gpt2, devlin2018bert}.
In particular, BERT~\cite{devlin2018bert} and its variants have enabled large improvements over the previous state of the art on a number of benchmark tasks~\cite{wang2019glue}. 

In this paper we take inspiration from BERT-style pretraining, specifically its use of masked reconstruction loss, and adapt the idea for speech recognition.
BERT is a bidirectional model that takes as input text that has had a certain percentage of randomly selected tokens masked, and attempts to reconstruct the masked text.  The idea is that a model that can predict the missing data should provide a good representation of the important content.
The same idea should hold for speech, but
there are some significant differences between text and speech signals.
In particular, the speech signal is continuous while text is discrete; and speech has much finer granularity than text, such that a single word typically spans a large sequence of contiguous frames.  
To handle these properties of speech, we take our second inspiration from recent work on speech data augmentation~\cite{Park_19a}, which applies masks to the input in both the time and frequency domains.
Thus, rather than randomly masking
a certain percentage of frames (as in BERT training),
we randomly mask some channels across all time steps of the input sequence, as well as contiguous segments in time.
We experiment with a range of choices for the number and width of masks, and find that for appropriate choices our BERT-style pretraining significantly improves over strong speech recognition baselines. 



\vspace{-.05in}
\section{Related work}
\label{s:related}
\vspace*{-1.5ex}

\karen{Kinda rewrote this section.  Took out the RBM paragraph, as I think it's sufficiently covered in the intro and it's not so central to the point of the paper, and some stuff about BERT that's covered in other sections.  Also removed the next couple of sentences as they were implying that recent work focuses on non-local RNN-based representations, but wav2vec is CNN-based and arguably a local window representation.  Finally, added some text about unsup learning in speech more broadly, and some text explicitly pointing out new stuff in this paper.}

Recent work has considered unsupervised learning for a variety of speech tasks.  Some of this work is explicitly aimed at a ``zero-speech'' setting where no or almost no labeled data is available at all (e.g.,~\cite{dunbar2019zero,Chorowski2019UnsupervisedSR,kamper2017segmental,harwath2016unsupervised}), where the focus is to learn phonetic or word-like units, or representations that can distinguish among such units.  Other work considers a variety of downstream supervised tasks, and some focuses explicitly on learning representations that generalize across tasks or across very different domains~\cite{pascual2019learning,chung2019unsupervised,hsu2017unsupervised,milde2018unspeech}.  This work uses a variety of training objectives, including autoencoder-based~\cite{Chorowski2019UnsupervisedSR} and language model-like~\cite{chung2019unsupervised}.

Specifically for our setting of unsupervised pre-training for supervised ASR, Schneider {\it et al.}~\cite{schneider2019wav2vec} and Pascual {\it et al.}~\cite{pascual2019learning} learn unsupervised convolutional network-based representations, and show that they improve the performance of ASR trained on smaller labeled data sets.  Their work relates to a number of other recent approaches for unsupervised representation learning~\cite{oord2018representation,hjelm2018learning} \karen{removed ``and their extensions'', and moved chung2019 to the prev paragraph as it's not MI-based and hasn't been used for ASR}
based on the idea of maximizing (a lower bound on) mutual
information (MI) between the current-time-step representation and 
future-time-step inputs (or shallow features of the inputs).
Such approaches use either convolutional or unidirectional architectures to extract
representations from audio, as their objective relies on the notion
of ``future'', which is not applicable for bidirectional models.  These methods obtain impressive results, but are not directly applicable to pre-training bidirectional RNNs, though they can in principle be stacked with bidirectional RNNs.
Concurrent work~\cite{anonymous2020vqwavvec} combines a mutual information-based approach with vector quantization for learning discrete representations, which are then used as input to BERT--an example of stacking a bidirectional model on top of a unidirectional MI-based one.  

Our work contrasts with prior work in several ways.  First, to the best of our knowledge our work is the first to pre-train bidirectional RNNs for direct use in a speech recognizer and to show improved recognition in this setting.  Besides the concurrent work of~\cite{anonymous2020vqwavvec}, we believe our work is also the first to use BERT-style masked reconstruction for representation learning for speech recognition.  In addition, we use continuous spectrogram-based input, which allows us to explore both time- and frequency-domain masking, and produces an overall much simpler method.  Finally, unlike other recent unsupervised pre-training approaches, we explicitly consider the problem of domain mismatch between the pre-training and fine-tuning data sets (see Section~\ref{sec:phone_libri}), and show that a simple adaptation layer can help address it.

\vspace{-.05in}
\section{Pre-training by masked reconstruction}
\label{s:method}
\vspace*{-1.5ex}

\begin{figure}[t]
    \centering
    \includegraphics[width=\linewidth,bb=0 20 780 745,clip]{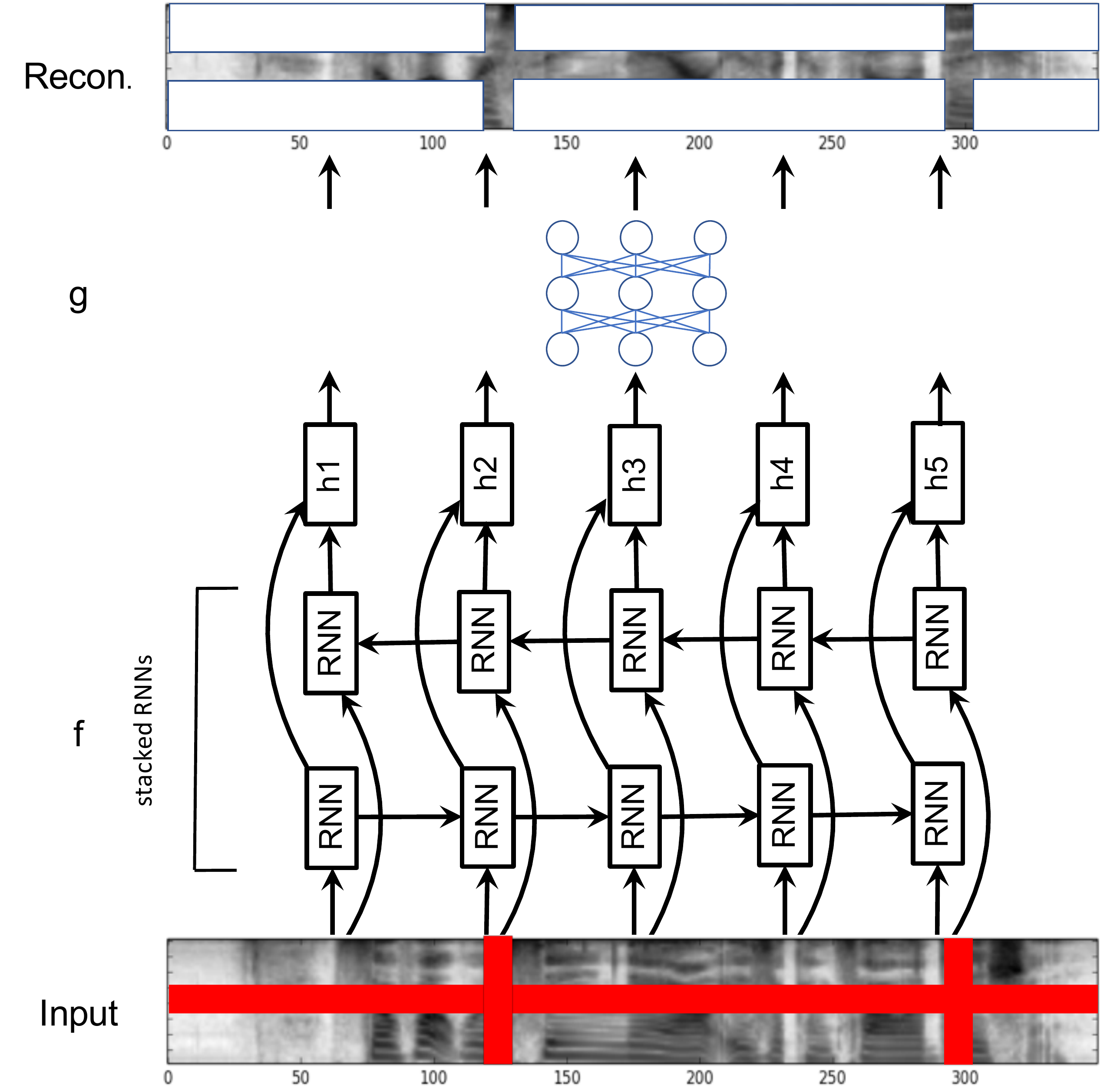}
    \vspace*{-2.5ex}
    \caption{Illustration of our masked reconstruction approach. \karen{changed ``loss'' to ``approach'' as this is not really illustrating the loss.  Minor esthetic suggestions for the fig:  remove the time axis labels in the output, remove the ``RNN''s from the encoder.}}
    \label{f:augmentation}
\vspace*{-1ex}
\end{figure}

The main idea of BERT training is to perturb the inputs by randomly masking tokens with some probability, and reconstruct the masked tokens at the output. \karen{removed a sentence}
Inspired by this idea, we perform representation learning for speech by masked reconstruction. Unlike the text domain where the inputs are discrete tokens, in the speech domain, the inputs are usually multi-dimensional feature vectors (e.g., energy in multiple frequency bands) in each frame, which are continuous and vary smoothly over time.  Moreover, the time span of each frame is typically
tens of milliseconds, much shorter than the span of the modeling unit in ASR.  Our approach adapts the idea of masked reconstruction to the speech domain.

Our approach can also be viewed as extending the data augmentation technique SpecAugment~\cite{Park_19a}, which was shown to be useful for supervised ASR, to unsupervised representation learning.  We begin with a spectrogram representation of the input utterance.  Viewing each input utterance as an image of dimension $D\times T$, where $D$ is the number of frequency bins and $T$ the number of frames, we adopt the spectral masking technique of~\cite{Park_19a} for masking the inputs:  We select $m_F$ segments of the input in the frequency axis with random locations, whose widths are drawn uniformly from $\{0, 1, \dots, n_F\}$, and similarly select $m_T$ segments in the time axis, with widths up to $n_T$, and set the selected pixels (time-frequency bins) to value $0$.  The intent is that masking in both frequency and time should encourage the network to exploit spatio-temporal patterns in the input.

Fig.~\ref{f:augmentation} illustrates our approach. 
Each input utterance $X$ is perturbed with a binary mask $M$ of the same dimensions as $X$ ($M_{td}=0$ if $X_{td}$ is being masked, for $t=1,\dots,T$ and $d=1,\dots,D$), and then
passed through a feature extractor $f$ consisting of several bidirectional recurrent neural layers \karen{removed LSTMs}
followed by a linear layer, to obtain a high level representation (features) for each frame.  Another (deep feedforward) network $g$ is then used to reconstruct the input from the features.  We measure the loss on the masked portion of the input:
\begin{gather*}
\calL(X, M; f, g) = \norm{ (1-M) \odot [X - g(f(M\odot X))] }_{Fro}^2
\end{gather*}
where $\odot$ denotes element-wise multiplication. 
Given a set of unlabeled utterances, we minimize the average of this reconstruction loss over the set.  After unsupervised pre-training, we retain the LSTM layers of $f$ and use them as initialization for supervised ASR training. 

To augment the unlabeled data, we also make use of the speed perturbation method of~\cite{Park_19a}, which performs linear interpolation along the time axis.
\karen{removed a phrase}
Besides the original data, we use two additional speed factors $0.9$ and $1.1$, effectively obtaining 3 times as much data for pre-training.

\karen{consider adding a point about acoustic domain adaptation, as this is the first paper that takes this into account for rep learning (I think)}

\vspace{-.05in}
\section{Experiments}
\label{s:expt}
\vspace*{-1.5ex}

\vspace{-.05in}
\subsection{Setup}
\label{s:setup}
\vspace*{-1.5ex}
We demonstrate our pre-training method using the
LibriSpeech~\cite{Panayotov_15a} and WSJ corpora\footnote{LDC catalog numbers LDC93S6B and LDC94S13B. \karen{shortened}}.
We explore a few settings with different amounts of data for
unsupervised pre-training and supervised fine-tuning. Supervised training
is always performed on WSJ,  with either the \emph{si84} partition (7040
utterances, 15 hours) or the
\emph{si284} partition (37.3K utterances, 80 hours) as the training set; the
\emph{dev93} partition (503 utterances) is used as development set, and
the \emph{eval92} partition (333 utterances) as the test set. The LibriSpeech
corpus, with a total of 960 hours of speech, is used for pre-training only.

The input consists of 40-dimensional log mel \karen{right?} filter bank energy (LFBE) features with a window size of 25ms and hop size
of 10ms, with per-speaker mean
 normalization for WSJ but not for LibriSpeech (we do not
 use any information beyond the audio of LibriSpeech). To speed
 up training, after data augmentation we stack every 3 consecutive frames.

We investigate the effect of pre-training on phone-based and character-based connectionist temporal classification (CTC) systems~\cite{Graves_06a}. 
The phone-based system uses a token set of 351 position-dependent phones, 
generated by the Kaldi \emph{s5} recipe~\cite{Povey_11a}. 
The character-based system uses 60 characters including the alphabet,
digits, and punctuation symbols. 
Acoustic model training is implemented with TensorFlow~\cite{Abadi_15a}; we use its beam search algorithm, with a beam size of 20, for
evaluating phone/character error rates.


Our acoustic model consists of 4 bidirectional LSTM
layers~\cite{HochreitSchmid97a} with 512 units in each direction. For
pre-training, the output feature space of $f(X)$ \karen{edited} has a dimensionality of 128. The
reconstruction network $g$ has two hidden layers of 1024 ReLU~\cite{NairHinton10a} units each. 
We use Adam~\cite{KingmaBa15a} as the optimizer for both pre-training and
fine-tuning, with initial learning rate tuned by grid search, mini-batch size $4$ for fine-tuning on \emph{si84} and $16$ for \emph{si284}, and maximum number of epochs 50.
We apply dropout~\cite{Srivas_14a} at all
layers, with rate tuned over $\{$0.0, 0.1, 0.2, 0.5$\}$.
We use the development set phone error rate (PER) at the end of each epoch as the criterion for hyperparameter search and early stopping.  The learning rate and
dropout are tuned once for the supervised baseline, and the resulting
values are used in all fine-tuning experiments.
For pre-training, optimization parameters are tuned to minimize the dev set
reconstruction loss, which happens within 15 epochs.
We set the maximum mask widths to $n_F=8$ and $n_T=16$, and tune the numbers of masks $m_F$ and $m_T$ based on development set ASR performance.

\vspace{-.1in}
\subsection{Phone-based: Pre-train on \emph{si284}}
\label{s:base}
\vspace*{-1.5ex}

We first pre-train the acoustic model on \emph{si284} and fine-tune
it on \emph{si84}, to investigate the effect of masking parameters
used in pre-training.
Note that in this setting, there is no domain difference between
pre-training and fine-tuning. The supervised baseline yields a dev PER of 18.52\%. \karen{I recommend reporting all numbers to a single decimal place}

\begin{table}[t]
\caption{Dev set \%PERs obtained by phone-based systems pre-trained on
  \emph{si284} and fine-tuned on \emph{si84}, 
  using different numbers of frequency masks ($m_F$) and time masks
  ($m_T$). The baseline PER without pre-training is 18.52\%.}
\label{t:tune_mfmt}
\vspace*{-1ex}
\centering
\begin{tabular}{@{}|c|c|c|c|c|@{}}
\hline
& $m_T=0$ & $m_T=1$ & $m_T=2$ & $m_T=3$ \\ \hline
$m_F=0$ & 18.33 & 18.51 & 17.83 & 18.20 \\
$m_F=1$ & 17.56 & 17.69 & \textbf{17.18} & 17.29 \\ 
$m_F=2$ & 17.29 & 17.53 & 17.47 & 17.40 \\
$m_F=3$ & 17.76 & 17.57 & 17.54 & 17.49 \\
\hline
\end{tabular}
\end{table}

Table~\ref{t:tune_mfmt} gives the dev set PERs after fine-tuning, with $m_F,
m_T \in \{0,1,2,3\}$. The case $m_F=m_T=0$ corresponds to reconstructing 
all
of the input spectrogram, which reduces to
the normal auto-encoder objective, and does not significantly \karen{right?} improve the acoustic
model.  This indicates that it is hard for the standard auto-encoder
approach to learn useful representations with this bidirectional
architecture, perhaps because given the full context, the reconstruction problem becomes too easy.
We also observe that it is important to have at least one frequency mask,
demonstrating the importance of exploring the joint time-frequency structure.
To verify the importance of masking segments rather than individual frames
or frequency bins, 
we pre-train another model where the total numbers of masked frames and frequency bins
are the same as those of our method using the best parameters
($m_F=1$, $m_F=2$),
but without constraining the masks to be contiguous; this model gives a worse dev PER of
17.61\%.



Based on the above results, we fix $m_F=1$ and $m_T=2$ for pre-training
phone-based systems. For this model pre-trained on \emph{si284}, we
fine-tune with different amounts of supervised data, with or without
augmenting the training set (using SpecAugment). The dev set PERs are given in
Table~\ref{t:fine-tune_phone} (second column).
We observe that pre-training is clearly helpful when the supervised set
is small (i.e., \emph{si84}).

\begin{table}[t]
\caption{Dev set \%PERs of phone-based systems fine-tuned
  with different amounts of supervised data, and initialized with different
  pre-trained models.}
\label{t:fine-tune_phone}
\vspace*{-1ex}
\centering
\begin{tabular}{@{}|l|r|r|r|r|@{}}
\hline
& Baseline & \caja{c}{c}{Pre-train\\ \emph{si284}} &
\caja{c}{c}{Pre-train\\ \emph{Libri.}\\ w/o LIN} & 
\caja{c}{c}{Pre-train\\ \emph{Libri.}\\ w/ LIN} \\ \hline
\emph{si84}                    & 18.52 & \textbf{17.18} & 17.61 &  17.31 \\
 + SpecAug                     & 16.83 & 15.56 & 15.64 &  \textbf{14.92} \\ 
\emph{si284}                  &   9.16 &   9.23 &   9.15 &  \textbf{ 8.50} \\
 + SpecAug                      &   7.98 &   8.21 &   8.19 &  \textbf{ 7.46} \\
\hline
\end{tabular}
\end{table}

\vspace{-.1in}
\subsection{Phone-based: Pre-train on LibriSpeech}
\label{sec:phone_libri}
\vspace*{-1.5ex}

We next explore how the amount and domain of the unlabeled data affect performance, \karen{edited} by pre-training on LibriSpeech
with 960 hours of speech.  For pre-training we use mini-batch size 128, and find that early stopping occurs after 7 epochs.
Since there is a domain difference between LibriSpeech and WSJ, we
also investigate the effect of domain adaptation for fine-tuning.  For
domain adaptation we use linear input
network (LIN,~\cite{Neto_95a,Yao_12a}), which inserts an additional linear layer
(initialized as the identity mapping) between the input and the pre-trained network, and only adapts this layer
and the softmax layer for the first 5 epochs of supervised training.

The dev set performance when pre-training on LibriSpeech is given in
Table~\ref{t:fine-tune_phone}, with or without LIN adaptation. We observe
that without LIN, the performance improvement tends to be smaller than
that of pre-training on \emph{si284}.  With LIN adaptation, we obtain consistently better
PERs, even when fine-tuning on \emph{si284}. Furthermore, the gains from
pre-training and SpecAugment are additive.

\vspace{-.1in}
\subsection{Character-based: Pre-train on LibriSpeech}
\vspace*{-1.5ex}
To study how pre-training interacts with
different modeling units, we repeat the above experiments for
character-based systems. We tune the masking parameters as before
(pre-train on \emph{si284} and fine-tune on \emph{si84}), and set $m_F=3$
and $m_T=2$ for pre-training on LibriSpeech.

\begin{table}[t]
\caption{Dev set \%CERs of character-based systems pre-trained on
  LibriSpeech, and fine-tuned with different amounts of supervised data.}
\label{t:fine-tune_char}
\vspace*{-1ex}
\centering
\begin{tabular}{@{}|l|r|r|r|r|@{}}
\hline
& Baseline &
\caja{c}{c}{Pre-train\\ \emph{Libri.}\\ w/o LIN} & 
\caja{c}{c}{Pre-train\\ \emph{Libri.}\\ w/ LIN} \\ \hline
\emph{si84}                    & 15.23  & 14.02 & \textbf{13.29}  \\
 + SpecAug                     & 12.98  & 12.26 & \textbf{11.70}  \\ 
\emph{si284}                  &   7.01  &   6.90 & \textbf{  6.48}  \\
 + SpecAug                      &   6.29  &   6.19 & \textbf{  5.61} \\
\hline
\end{tabular}
\end{table}

\begin{figure}[t]
\centering
\begin{tabular}{@{}c@{\hspace{0\linewidth}}c@{}}
\includegraphics[width=0.50\linewidth,bb=0 0 430 320,clip]{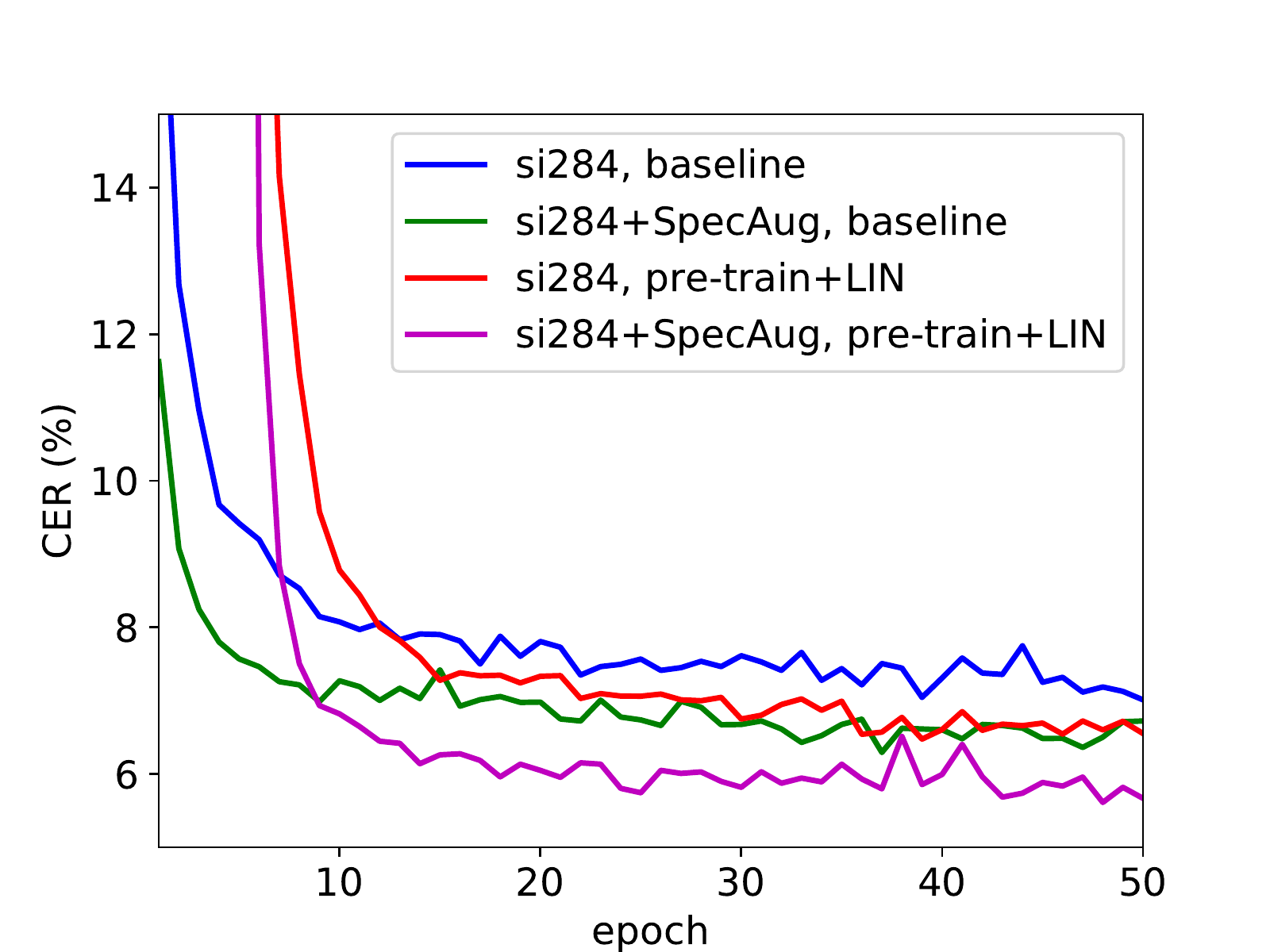} & 
\includegraphics[width=0.50\linewidth,bb=0 0 430 320,clip]{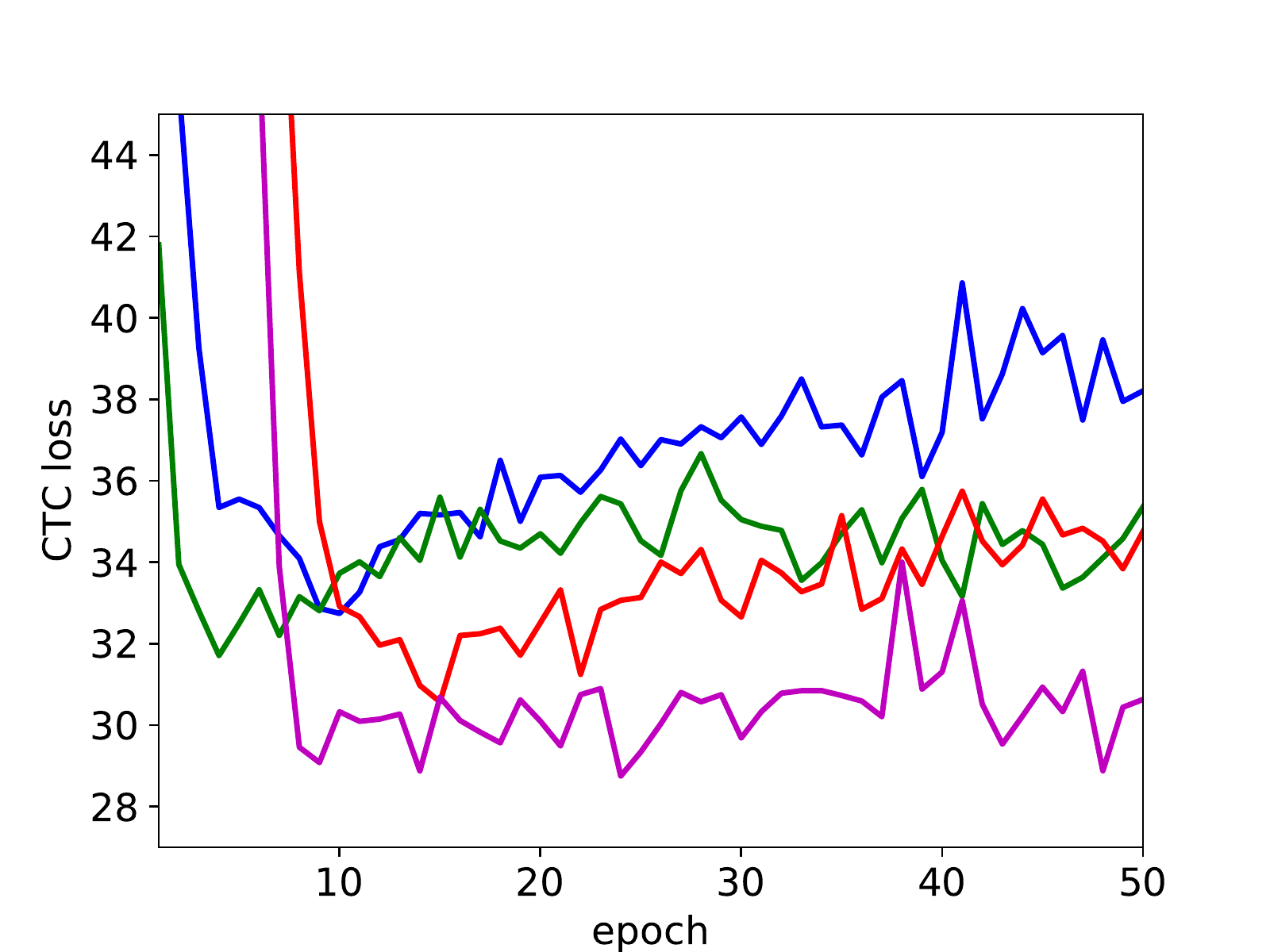}
\end{tabular}
\vspace*{-2ex}
\caption{Dev set learning curves (\%CER and CTC loss) of different systems pre-trained on \emph{LibriSpeech}. The first 5 epochs of
  fine-tuning update only the LIN and softmax layers.}
\label{f:learn_curves}
\vspace*{-1ex}
\end{figure}

The dev set performance of pre-trained character-based systems is given in
Table~\ref{t:fine-tune_char}. 
The observations are consistent with those on the phone-based systems.
Fig.~\ref{f:learn_curves} shows learning curves of the CTC
systems in terms of both CER and the average CTC loss over the dev set,
with or without SpecAugment. We see that, although the two criteria
(CER and loss) do not synchronize completely, the pre-trained systems are
advantageous in terms of both. Note that all four models are trained with the same
optimization parameters, and the loss curves with pre-training generally
show less overfitting.

\begin{table}[t]
\caption{\%WERs obtained by character-based CTC systems on the test set.
Pre-training is done on \emph{LibriSpeech}.}
\label{t:wers}
\centering
\begin{tabular}{@{}|l|r|@{}}
\hline
Method & WER\\ \hline
EESEN~\cite{Miao_15a} (extended tri-gram) & 7.34 \\ \hline
\emph{si284}                                        & 7.69  \\
\emph{si284}   + SpecAug                     & 7.44  \\ 
\emph{si284}   + pre-train + LIN               &  6.66 \\
\emph{si284}   + SpecAug + pre-train + LIN &  \textbf{6.33} \\
\hline
\end{tabular}
\end{table}

Finally, we evaluate word error rates (WERs) for the above character-based systems using the
WFST-based framework of Miao {\it et al.}~\cite{Miao_15a}, with the extended 4-gram language
model built by the Kaldi recipe. After composing
the decoding (TLG) \karen{remove ``(TLG)'' entirely?} graph, we perform beam search using Kaldi's
\texttt{decode-faster} with beam size 20 and acoustic model scale 
tuned on the dev set.
Test set WERs are given in Table~\ref{t:wers}.  For reference, EESEN's
character-based system obtains a test WER of 7.34\% with a different
language model, when trained on \emph{si284}. \karen{might need to justify why we include this baseline if it uses a different LM.} Our results show
that the more accurate pre-trained acoustic models also give improved word-level decodings with a language model.


\vspace{-.1in}
\section{Conclusions}
\label{s:conclusion}
\vspace*{-1.5ex}

This work demonstrates that pre-training by masked reconstruction
leads to consistent performance improvement for CTC-based
ASR. Some questions remain open.
We have chosen different masking parameters for
pre-training the phone-based and character-based systems, by
tuning on the development set; it would be good to have a more efficient way of choosing these hyperparameters.  In addition, a thorough comparison is needed with other recent work on representation learning approaches~\cite{schneider2019wav2vec,chung2019unsupervised,pascual2019learning,anonymous2020vqwavvec} to separate the effects of model type versus pre-training approach;
this is not trivial as it is not straightforward
to extend these prior approaches to bidirectional recurrent models.
\karen{removed the last sentence as it seemed weak to me.  we could maybe come up with a better version of the same sentiment}

\vspace{-.1in}
\section{Acknowledgements}
\vspace*{-1.5ex}
The authors would like to thank Yang Chen for helpful discussions, and Amazon for providing computational resources.
This material is based upon work supported by the Air Force Office of Scientific Research under award number FA9550-18-1-0166 and by NSF award number 1816627.

\vspace{-.1in}
\bibliographystyle{IEEEbib}
\bibliography{refs}
\end{document}